\documentclass{JHEP3}
\usepackage{epsfig}
\usepackage{graphicx}
 \def\be{\begin{equation}}
\def\ee{\end{equation}}
\def\bea{\begin{eqnarray}}
\def\eea{\end{eqnarray}}
 
\def\lesssim{\mathrel{\hbox{\rlap{\hbox{\lower4pt\hbox{$\sim$}}}\hbox{$<$}}}}
\def\gtrsim{\mathrel{\hbox{\rlap{\hbox{\lower4pt\hbox{$\sim$}}}\hbox{$>$}}}}

\title{Fast-Roll Inflation}

\author{Andrei Linde\\
    Department of Physics, Stanford University, Stanford, CA 94305,
USA\\
    E-mail: \email{linde@physics.stanford.edu}\\
    http://physics.stanford.edu/linde}
\received{October 21, 2001}       
 \preprint{SU-ITP-01/45\\  ~hep-th/0110195\\ October 21, 2001}

\abstract{We  show that in the simplest theories of spontaneous symmetry breaking one can have a stage  of a {\it fast-roll inflation}. In this regime the standard slow-roll condition $|m^2| \ll H^2$ is violated. Nevertheless, this stage can be rather long if $|m|$ is sufficiently small. Fast-roll inflation can be useful for generating proper initial conditions for the subsequent stage of slow-roll inflation in the very early universe. It may also be responsible  for the present stage of accelerated expansion of the universe.  We also make two observations of a more general nature. First of all, the universe after a long stage of inflation (either slow-roll or fast-roll) cannot reach anti-de Sitter regime even if the cosmological constant is negative. Secondly, the theories with the potentials with a ``stable'' minimum at $V(\phi)<0$ in the cosmological background exhibit the same instability as the theories with potentials unbounded from below. This instability leads to the development of singularity with the properties practically independent of $V(\phi)$. However, the development of the instability in some cases may be so slow that the theories with the potentials unbounded from below can describe the present stage of  cosmic acceleration even if this acceleration occurs due to the fast-roll inflation.}

\keywords{eld.pbr.ctg.sgm}

\begin{document}

\section{Introduction} 

Most of the realistic versions of inflationary cosmology such as new inflation \cite{New}, chaotic inflation \cite{chaot} and hybrid inflation \cite{hybrid}  are based on the assumption that the scalar field $\phi$ driving inflation satisfies certain slow-roll conditions. One of the simplest conditions is $|m^2| \ll H^2$, where $m$ is the mass of the inflaton field $\phi$ and $H$ is the Hubble constant \cite{book}. If one uses the system of units $M_p = 1$ (where $M_p = 2.4 \times 10^{18}$ GeV $=(8\pi G)^{-1/2})$, one has $H^2 = V/3$ during inflation, and the condition $|m^2| \ll H^2$ can be expressed in a simple form $\eta = {|V''|\over |V|} \ll 1$. The second slow-roll condition is  $\epsilon = {1\over 2} \left({V'\over V}\right)^2 \ll 1$ \cite{LythLiddle}. 

The slow-roll conditions serve two purposes: They make the total expansion of the universe during the stage of inflation very large,  and they ensure that the spectrum of adiabatic density perturbations produced during inflation  is almost scale-independent. Density perturbations are produced due to quantum effects during inflation \cite{pert}. They are playing a critical role in the subsequent process of formation of the large-scale structure of the universe \cite{book,LythLiddle}. For $\eta \ll 1$ and $\nu \ll 1$, the deviation from scale-independence (flatness) of the spectrum is characterized by the factor $n-1 \approx  2 \eta -6\epsilon$ \cite{LythLiddle}. If the slow-roll conditions are satisfied, one has $n \approx 1$. Recent observations of anisotropy of the cosmic microwave background radiation suggest that $|n-1| \lesssim 0.1$ \cite{CMB}.

However, there are many theories where one has $|m^2| = O(H^2)$. For example, if the field $\phi$ nonminimally couples to gravity, it may acquire a correction to the mass squared $\Delta m^2 = \xi R$. Here $R$ is the curvature scalar, $R = 12 H^2$ during inflation. For the popular choice $\xi = 1/6$ (conformal coupling) one finds  $\Delta m^2 = 2H^2$. A similar situation appears in N=1 supergravity, where the scalar fields quite often acquire corrections $\Delta m^2 =O(H^2)$ \cite{dinefisch}.

Recently a closely related issue has been studied in the context of N=8, N=4, and N=2 gauged supergravity. It was found that in all known models of this type every extremum of the  effective potential of the scalar field with  $V(\phi) > 0$ corresponds either to a maximum or to a saddle point. The curvature $V''$ of the effective potential in the direction of the steepest descent in all of these models is equal to $-2V$. This corresponds to the tachyonic (i.e. negative) mass squared $V'' = -6H^2$ \cite{KLPS}.

Thus we see that there are many theories where $|m^2| = O(H^2)$, i.e. the slow-roll condition $\eta = {|V''|\over |V|} \ll 1$ is violated. Does this mean that such theories cannot describe inflation/acceleration of the universe? 

This question was addressed in \cite{KLPS}, where a simple model of the fast-roll inflation was presented. In this paper we will briefly summarize  the main idea of the fast-roll inflation and discuss possible implications of the new scenario.
Fast-roll inflation can be useful for generating proper initial conditions for the subsequent stage of slow-roll inflation in the very early universe. It may also be responsible  for the present stage of accelerated expansion of the universe.  

In the last case, one make consider four distinct possibilities: The potential for the scalar field may have a minimum at $V(\phi) > 0$, at $V(\phi) = 0$, or at $V(\phi) < 0$, or it may be unbounded from below.  Naively, one could expect that if the potential $V(\phi)$ has a minimum,  the field $\phi$ eventually rolls to the minimum of $V(\phi)$ and stays there: If $V(\phi)$ has a minimum at $V>0$, the universe will eventually behave as de Sitter space; the minimum with $V=0$ will correspond to Minkowski regime, whereas in the case when the minimum occurs at $V<0$, we will wind up in anti-de Sitter space with a negative cosmological constant. 

However, we will show that when one considers this issue in the cosmological background, the behavior of the universe with $V(\phi) <0$ is more complicated.
First of all, the universe after a long stage of inflation (either slow-roll or fast-roll) cannot reach anti-de Sitter regime even if the cosmological constant is negative. This is a simple consequence of the Friedmann equation $H^2 = {\rho(\phi)/3}$ ~for the flat universe. The second observation is related to the first one but is a little bit less trivial: The theories with potentials having a ``stable'' minimum at $V(\phi)<0$ exhibit the same kind of instability in the cosmological background  as the theories with potentials unbounded from below. This instability leads to the development of singularity with the properties practically independent of $V(\phi)$. Thus, the existence of the minimum does not yet guarantee safety. On the other hand, the development of the instability in some cases may be so slow that the theories with the potentials unbounded from below, as well as the potentials with the minimum at $V(\phi) < 0$, can describe the present stage of  cosmic acceleration even if this acceleration occurs due to the fast-roll inflation.

\section{ Dynamics of spontaneous symmetry breaking and fast-roll inflation}

Consider  a theory of a scalar field $\phi$ with potential $V(\phi)$ and energy density $\rho(\phi) = V(\phi) + \dot\phi^2/2 + (\partial_i \phi)^2/2$. The general Friedmann equation in units $M_p = 1$ (which we are going to use in the main part of this paper)  looks as follows:
\bea\label{freedgen}
H^2 +  {k\over a^2} = \left({\dot a\over a}\right)^2 +  {k\over a^2}= {\rho(\phi)\over 3} \ . 
\eea
Here $k = \pm 1,0$ for a closed, open or flat universe correspondingly. In this paper we will assume that the universe is flat, which is a very good approximation once inflation begins. In this case
\bea\label{freed}
H^2 = \left({\dot a\over a}\right)^2  = {\rho(\phi)\over 3} \ .  
\eea

Suppose that the potential $V(\phi)$ has a maximum at $\phi = 0$. Usually one can represent any such potential as
\bea\label{simplepot}
V(\phi) = V_0 - {m^2\phi^2\over 2} \  
\eea
in the vicinity of the maximum. The last term implies that the field $\phi$ near the top of the potential corresponds to the tachyonic (unstable) mode with the mass squared $V'' = -m^2$.

We will assume for simplicity that this expression for $V(\phi)$ is valid at least for $|\phi| \leq \phi_*$, where $V(\phi_*) = V_0/2$, i.e. \bea
\phi_*  = {\sqrt {V_0} \over m} \  .
\eea
What is so special about $\phi_*$?  Suppose the field falls down from $\phi \ll \phi_0$ to $\phi_*$ during the time $\Delta t$ with a small initial kinetic and gradient energy $\dot\phi^2/2, (\partial_i \phi)^2/2 \ll V_0$. If the motion of the field slows down because of the rapid expansion of the universe, the energy density $\rho(\phi) \approx V(\phi)$ decreases 2 times when the field reaches $\phi = \phi_*$, and the Hubble constant decreases only by a factor of $1/\sqrt 2$. If the field does not slow down, $\rho$ remains approximately equal to $V_0$. Thus in both cases $H$ remains approximately constant, so that one can use the following estimate for the total growth of the universe during inflation:
\bea 
{a(t_*)\over a_0} \approx   e^{H  t_*} \  ,
\eea
where $H^2 \approx V_0/3$, and $t_*$ is the time when the field $\phi$ reaches $\phi_*$. 

In the theories with $\phi_* \gg 1$ the universe may expand nearly exponentially even for $\phi > \phi_*$, as in the simplest chaotic inflation models \cite{chaot}. In this paper we will restrict ourselves to the  models with $\phi \lesssim 1$. In this case inflation, if it ever begins, ends at $\phi \sim \phi_*$.

If the field $\phi$  is sufficiently homogeneous, its
motion  in the theory (\ref{simplepot}) for $H = const$ can be described by equation \cite{book}
\bea\label{genequation}
\ddot\phi + 3H\dot\phi = - V' = m^2\phi \ .
\eea
One may look for solutions of Eq. (\ref{genequation}) in the form $\phi = \phi_0 e^{i\omega t}$. This yields
\cite{book}
\bea\label{genequation2}
\omega^2 - 3iH\omega + m^2 = O\ ,
\eea
which gives
\bea    
\omega =i \left({3H\over 2} \pm \sqrt{{9H^2\over 4} +m^2} \right). 
\eea
The solution with the sign + corresponds to the exponentially decreasing field, which rapidly disappears, whereas the solution with the sign - corresponds to the exponentially growing field $\phi$:  
\bea \label{phit}   
\phi(t) =\phi_0 \exp \left [\left(Ht\cdot F(m^2/H^2) \right)\right] , 
\eea
where 
\bea    
F({m^2/H^2}) = \sqrt{{9\over 4} + {m^2\over H^2}} - {3\over 2} \ . 
\eea
Equation (\ref{phit}) allows us to give a simple estimate for the total expansion of the universe during the period of inflation:
\bea \label{at}   
e^{Ht_*} =\left({\phi_*\over \phi_0}\right)^ {1/F} . 
\eea

There are two important limiting cases for which these results can be written more explicitly.  For $m \ll H$ (slow-roll inflation) one has
\bea  \label{slowinfl}  
F({m^2/H^2}) =  {m^2 \over 3H^2 } \ , 
\eea 
\bea    
\phi =\phi_0 \ \exp{m^2t\over 3H}\ , 
\eea
and 
\bea \label{atslow}   
e^{Ht_*} =\left({\phi_*\over \phi_0}\right)^{3H^2/m^2} =  \left({\phi_*\over\phi_0}\right)^{V/|V''|}. 
\eea
The last result once again demonstrates the importance of the slow-roll condition $m^2 \ll H^2$: The smaller is the ratio $m^2/H^2$, the greater is the total expansion of the universe during the period of inflation.

Meanwhile, for $m \gg H$  (very fast roll) one has
\bea  \label{noinfl}  
F({m^2/H^2}) =  {m \over H } \ , 
\eea
\bea    
\phi =\phi_0\  e^{Ht}\ , 
\eea
and
\bea \label{atveryfast}   
e^{Ht_*} =\left({\phi_*\over \phi_0}\right)^ {H/m} . 
\eea

For the theories with $m^2 = O(H^2)$ the knowledge of these two limiting cases is instructive but insufficient. Fortunately, one can easily find the numerical values of the function $F^{-1}(x)$, which enters Eq. (\ref{at}), for all values of $x = m^2/H^2$. In particular, $F^{-1}(1) = 3.3$, $F^{-1}(6) = 0.73$, $F^{-1}(12) = 0.44$. We will use these results shortly.

But before going any further we must specify initial conditions. Formally, one can take $\phi_0$ as small as one wants, and get an indefinitely long stage of inflation. So what prevents us from taking $\phi_0 = 0$ and  having inflation during spontaneous symmetry breaking in the electroweak theory?

The answer is that $\phi_0$ in Eqs. (\ref{phit}) and (\ref{at}) cannot be taken smaller than the level of quantum fluctuations with momenta $k<m$, since such fluctuations also experience exponential growth, even in the absence of a homogeneous field $\phi_0$. For example, for $m\gg H$ one has
\bea \label{tachk}
\delta\phi_k(t) \sim \delta\phi_k(0)e^{\sqrt{m^2-k^2}\, t} .
\eea
Typically,  the growth of these quantum fluctuations, rather than the growth of the  homogeneous mode of the field $\phi$, is responsible for spontaneous symmetry breaking \cite{FKL}.

A typical initial amplitude of all quantum fluctuations with $k < m$ participating in the exponential growth of the field $\phi$ depends on the way one prepares initial conditions for inflation/spontaneous symmetry breaking. To get an idea of a possible amplitude of initial quantum fluctuations, we will follow \cite{FKL} and assume that 
 the mode functions describing quantum fluctuations in the symmetric phase $\phi=0$ at the moment close to  $t = 0$ are the same as for a massless field,
$\phi_k ={1 \over \sqrt{2 k}}e^{-ikt +i{\vec k \vec
x}}$. Then at $t = 0$ we `turn on' the term $-m^2\phi^2/2$
corresponding to the negative mass squared $-m^2$. The modes with
$k = |{\vec k}|  < m$ grow  exponentially. 
Initial dispersion of all growing fluctuations with $k < m$ was given by 
 \begin{equation}
\langle \delta\phi^2 \rangle
 =  \int\limits_0^{m}  {  dk^2  \over 8\pi^2 } = {m^2\over 8 \pi^2} \ ,
\label{aBBB}
\end{equation}
and the average initial amplitude of all fluctuations with $k < m$ was given by $\delta\phi \sim  {m/2 \pi }$ \cite{FKL}.
However, this  is a slight overestimate. For example, fluctuations with $k = m/2$ grow with a somewhat smaller speed than the fluctuations with $k = m/4$. Over a large time interval this small difference becomes more and more important. Consequently, if the field falls down during the time much greater than $m^{-1}$ (and this is the regime where one may get inflation), the main contribution to the growing field distribution is given by fluctuations with $k \ll m$. The initial amplitude of such fluctuations will be somewhat smaller than ${m/2 \pi }$. For simplicity, we will assume that the smallest possible effective value of $\phi_0$ in our equations is $m/C$ with $C = O(10)$. Numerical simulations performed in \cite{FKL} suggest that $C$ is somewhat greater than 10. Precise details will not be very important for our estimates, where we will simply replace $\phi_0$ by $m/10$ to find  the maximal possible value of the expansion factor $e^{Ht_*}$ for a given model:
\bea \label{atmax}   
e^{Ht_*}\sim \left({10\phi_*\over m}\right)^ {1/F} . 
\eea

Now we are ready to estimate the maximal possible value of the expansion factor $e^{Ht_*}$ in various models.
In the theories with the potentials (\ref{simplepot}) with $m = O(H)$  one has  $\phi_* \sim M_p = 1$, so that 
\bea \label{atmax2}   
e^{Ht_*}\sim \left({10\over m}\right)^ {1/F} . 
\eea
Consider now the simplest case  $m = H$. In this case one has $F^{-1}(1) =3.3$. In the theories with $m \sim H$ one has  $\phi_* \sim M_p = 1$, so for $\phi_0 \sim 10 m^{-1}$ (i.e. for the initial value of the field provided by quantum fluctuations \cite{FKL})  one has 
\bea \label{smallm}
e^{Ht_*} \sim  \left({10\over m}\right)^{3.3}  \ .
\eea
Clearly, for $m \ll 1$ (i.e. for $m \ll M_p$) this number can be quite significant.

To be more specific, consider an extreme possibility that such models are responsible for the present stage of accelerated expansion of the universe with the Hubble constant $H \sim 10^{-60}$ (i.e. $m = H \sim 10^{-60}M_p$) \cite{supernova,CMB}. Then inflation in an unstable state close to the maximum of the potential in such a theory can lead to expansion of the universe by a factor that can be as large as 
\bea
e^{Ht_*}\sim \left({10^{61}}\right)^{3.3} \sim 10^{200} \sim e^{460}~.
\eea
This is more than sufficient to explain the observed single e-folding of accelerated expansion of the universe at the present epoch.

Meanwhile if one takes $m \sim 10^2$ GeV $\sim 10^{-16} M_p$, which corresponds to the electroweak scale, one can obtain fast-roll inflation by a factor of 
\bea
e^{Ht_*} \sim  \left({10^{17}}\right)^{3.3} \sim 10^{56}\sim e^{130}~.
\eea
Let us take $m = 10^{-1}$, which corresponds to $m \sim 2\times 10^{17}$ GeV. In this case one finds
\bea\label{01}
e^{Ht_*} \sim  \left({10^{2}}\right)^{3.3} \sim 10^{6.6}\sim e^{15}~.
\eea
Even in the limiting case $m = H = 1$ we can have a short stage of inflation,
\bea\label{01a}
e^{Ht_*}  \sim 10^{3.3} \sim e^{7.5}~.
\eea

Efficiency of the fast-roll inflation rapidly decreases when one considers the regime with $H \ll m$, or when $m$ approaches the Planck mass scale $M_p = 1$. An interesting example is provided by gauged N=8 supergravity, where, as we already mentioned, $|V''| = 2V$, i.e. $m^2 = 6H^2$ \cite{KLPS}. In terms of our  potential $V(\phi) = V_0 - {m^2\phi^2\over 2}$ this implies that the point $\phi_*$, which corresponds to $V(\phi_*) = V_0/2$, is given by $\phi_* = 1/\sqrt 2$. In dimensional units this is equivalent to having $\phi_* \sim 1.7\times 10^{18}$ GeV.   In this model one has $F(m^2/H^2) = F(6) = 1.37$, and $1/F = 0.73$.
Then, for $m = \sqrt 6 \,H \sim 2\times 10^{-60}M_p$ one finds
\bea
e^{Ht_*} \sim \left(5\  10^{60}\right)^{0.73} \sim {10^{44}} \sim e^{100}  \ .
\eea
Thus, fast-roll inflation in N=8 gauged supergravity could be responsible for up to 100 e-folds of exponential expansion of the universe with the Hubble constant similar to its present value $H \sim  10^{-60}M_p$. 

On the other hand, for $m = \sqrt 6 H \sim   10^2$ GeV one has
\bea
e^{Ht_*} \sim  {10^{13}} \sim e^{28}  \ .
\eea 

One should note that all known  potentials of N=8 gauged supergravity with an extremum at $V>0$ are unbounded from below, see e.g.  \cite{Hull,Ahn:2001by} and \cite{KLPS}, so one needs to find some other potentials  to describe inflation in the early universe. However, it is quite interesting that if the energy scale corresponding to such potentials is extremely small, $m \sim  10^{-60}M_p$, then we can safely live near the top of the effective potential for hundreds of billions of years and experience exponential expansion... until the big crunch \cite{KLPS}. 

To give a different (and safe) example, one may consider a simplest model of inflation during the process of spontaneous symmetry breaking in the theory with the  potential
\bea \label{SSB}
V(\phi) = {\lambda\over 4} (\phi^2-v^2)^2 = -{1\over 2} m^2\phi^2 + {m^2 \over 4 v^2} \phi^4 +{1\over 4} m^2 v^2       \ .
\eea 
Here $m^2 = \lambda v^2$, and $\phi = v$ corresponds to the minimum of $V(\phi)$ with symmetry breaking. The Hubble constant at $\phi = 0$ in this model is given by $H^2 = {m^2 v^2\over 12}$, so that $F(m^2/H^2) = F(12/v^2)$. 

Inflation in this model  was considered a long ago in \cite{chaot2}; this was the second realization of chaotic inflation scenario \cite{chaot}. However, at that time we concentrated on the slow-roll regime with $m^2 \ll H^2$, $v \gg 1$. Now we have a different goal. Since we know that in many supergravity-inspired models we may have a natural scale for spontaneous symmetry breaking $v = O(1)$, we are going to find out whether such models may also lead to a stage of inflation.

One can easily check that the fast-roll inflation in the model (\ref{SSB}) with $m = O(H)$ occurs for $\phi \lesssim v/2$.  Assuming, e.g., $v = 1$ one finds $m^2 = 12 H^2$ and $F^{-1}(12) = 0.44$. For $H \sim 10^{-60}M_p$ one finds that the fast-roll inflation in this case can lead to expansion of the universe by a factor of
\bea\label{ssninfl}
e^{Ht_*} \sim {10^{26}} \sim e^{60}  \ ,
\eea
whereas for $m \sim 100$ GeV, one finds 
\bea\label{modulipr}
e^{Ht_*} \sim 10^{7}\ .
\eea 

\
 
\section{Possible implications of the fast-roll inflation}
\subsection{\label{now}Fast-roll inflation and non-eternal quintessence} 
One of the problems of the modern cosmology is to find a proper mechanism which drives the present stage of the accelerated expansion/inflation of the universe \cite{CMB,supernova}. The present inflationary stage may be driven either by the constant positive vacuum energy $V_0 \sim 10^{-120} M_p^4$, or by a scalar field with the potential which leads to a quasi-inflationary stage and which is equal to $V(\phi) \sim 10^{-120} M_p^4$  at the present moment. It is often assumed that $V(\phi)$ very slowly decreases at $\phi \to \infty$ \cite{quint}. In such models an accelerated expansion of the universe continues for an indefinitely long time (``eternal quintessence''). However, recently it was argued that it may be difficult to describe eternal quintessence in the context of M-theory \cite{Hellerman:2001yi}, and several suggestions have been made how one could make the present stage of acceleration of the universe long but finite, see e.g.  \cite{Halyo:2001fb,Kallosh:2001tm}. We believe that one of the easiest ways to do so is to use inflation in the simplest model of spontaneous symmetry breaking  (\ref{SSB}). If the parameter $v$ in this model is very large, $v \gg 1$, we will have the standard slow-roll inflation. For $v = 1$ we have a very long stage of the fast-roll inflation, $e^{Ht_*}\sim   e^{60}$, see Eq. (\ref{ssninfl}). Meanwhile the only thing we need is to explain the observed single e-fold of the accelerated expansion. Eqs. (\ref{atveryfast}) and (\ref{atmax2}) imply that one can achieve it even if $H \sim 10^{-2}m$, for $v \gtrsim 3\times 10^{-2} M_p  \sim 10^{17}$ GeV. 
In other words,  one can have accelerated expansion in the simplest model of spontaneous symmetry breaking with $H \sim 10^{-60}M_p$ even if the amplitude of spontaneous symmetry breaking $v$ is thirty times smaller than the Planck scale.  In this regime  the slow-roll condition $|m^2|\ll H^2$ is violated by 4 orders of magnitude!

Moreover, to describe the present stage of the accelerated expansion one can use even the simplest potential $V(\phi) = V_0 - {m^2\phi^2\over 2}$, as in Eq. (\ref{simplepot}). The potentials of that type appear in $N = 8$ supergravity, see e.g. \cite{Hull,Ahn:2001by,KLPS}. This potential is unbounded from below, but since this danger will reveal itself many billions of years in the future, one should not discard such models. 

In this respect I would like to make two observations. Both of them are rather simple, but I believe that they deserve some attention.

\subsubsection{Long stage of inflation cannot lead to the universe dominated by  $\Lambda < 0$.}
One of the aspects of the cosmological constant problem is the choice of sign of the vacuum energy $\Lambda$. In general, for the open universe case
one may have a solution to the Friedmann equation $H^2 -a^{-2} = \Lambda/3$ for $\Lambda <0$: ~ $a(t) = \sqrt{3\over |\Lambda|}~\sin \left(\sqrt{|\Lambda|\over 3}\, t\right)$. This is a specific section of anti de Sitter space that is so popular in M-theory and brane cosmology. Thus one may wonder whether it is possible to achieve the anti-de Sitter regime in the realistic cosmological models.  

The answer to this question is very simple, though perhaps somewhat unexpected: One cannot reach the anti-de Sitter regime after a long stage of inflation. In other words,  {\it one of the predictions of inflationary cosmology is that we cannot wind up in AdS space.} 

Indeed, after a long stage of inflation in the early universe, the universe becomes almost exactly flat. This effectively means that the term $k\,a^{-2}$ with $k = \pm 1,0$ can be omitted in the general Friedmann equation (\ref{freedgen}). The Friedmann equation $H^2   = \rho/3$ describing flat universe  (\ref{freed}) does not have any solutions with $\rho(\phi)< 0$. Therefore AdS universe with the energy density dominated by the  negative  cosmological constant cannot appear after inflation, unless one considers open inflation with $\Omega <1$. Thus, one of the consequences of the standard inflationary prediction $\Omega = 1$ is that our universe cannot reach the AdS regime. This observation is rather trivial, but the existence of relation between inflation in the early universe and the energy density at the present epoch is quite interesting.  It may be important to keep it in mind when discussing the cosmological constant problem:  {\it Because of inflation, we cannot live in the universe dominated by the negative cosmological constant.}

\subsubsection{The potentials with minima at $V(\phi)<0$ lead to a singular behavior of the field $\phi$, just like the potentials unbounded from below.}

 The previous comment implies  that, from the point of view of cosmology, the potentials that are bounded from below but have a minimum at $V(\phi) <0$ are not much better than  the potentials  unbounded from below. Indeed, the worst thing about the potentials unbounded from below is that the field $\phi$ can fall down to indefinitely large negative values of $V(\phi)$ causing some kind of singular behavior. But exactly the same (though for a slightly different reason) happens in any theory with the potential having a minimum with $V(\phi)<0$. According to the Friedmann equation $H^2 = \rho/3$, a homogeneous field $\phi$ cannot relax near the minimum with $V(\phi)<0$, it must move so that $\rho(\phi) = V+ \dot\phi^2/2$ remains non-negative. On the other hand, the kinetic energy of the field $\phi$ can only decrease during the stage of expansion. Once the universe reaches the regime $\rho(\phi) = V+ \dot\phi^2/2=0$, the Hubble constant vanishes, i.e. the universe reaches its maximal size and stops expanding. But this state is unstable; the universe bounces back, the Hubble constant becomes negative,  and the universe collapses, see e.g. \cite{Krauss:1999br,Kaloper:1999tt}. 

As we will see now, the cosmological singularity  involves the singular behavior of metric and of the scalar field. This behavior occurs  for the  potentials  bounded and unbounded from below, once the scalar field may reach  the regime with $V<0$.
This is a rather counter-intuitive conclusion which could not be reached if we would neglect the back-reaction of the scalar field on the expansion of the universe.

Here we present the results of the numerical investigation of the behavior  of the scalar field $\phi(t)$ and the scale factor $a(t)$ in two models. The first one is the model with $V(\phi) = V_0 - {m^2\phi^2\over 2}$, with the potential unbounded from below, see Fig. 1.  
As we see, in the beginning the universe experiences a stage of inflation.  At this stage the ``friction term'' $3H\dot\phi$ in the equation of motion of the scalar field becomes negative, which speeds up the growth of the field $\phi$ and leads to a rapid collapse of the universe. Then the field $\phi$ grows, the potential becomes negative, and soon after that the universe begins to collapse.

\begin{figure}[h!]
\leavevmode
\centering \epsfysize=4cm
\includegraphics[scale=0.56]{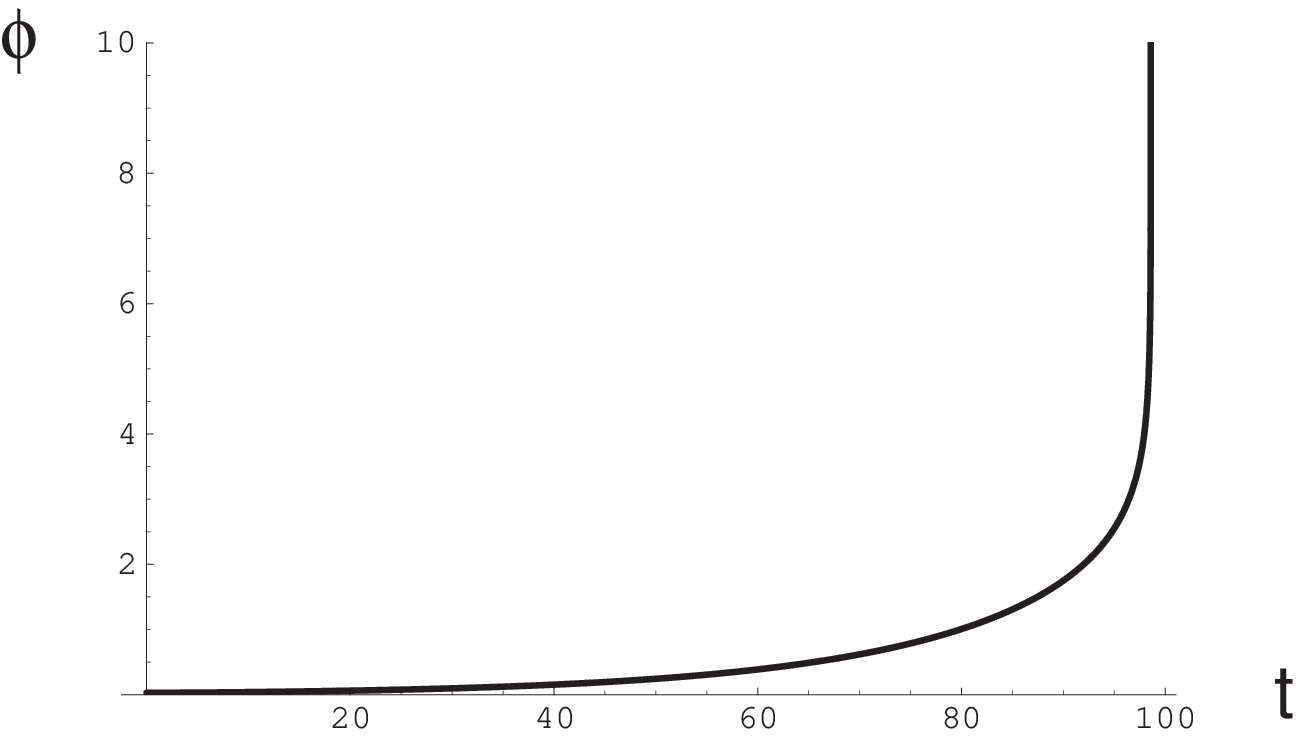}~~~
 \includegraphics[scale=0.56]{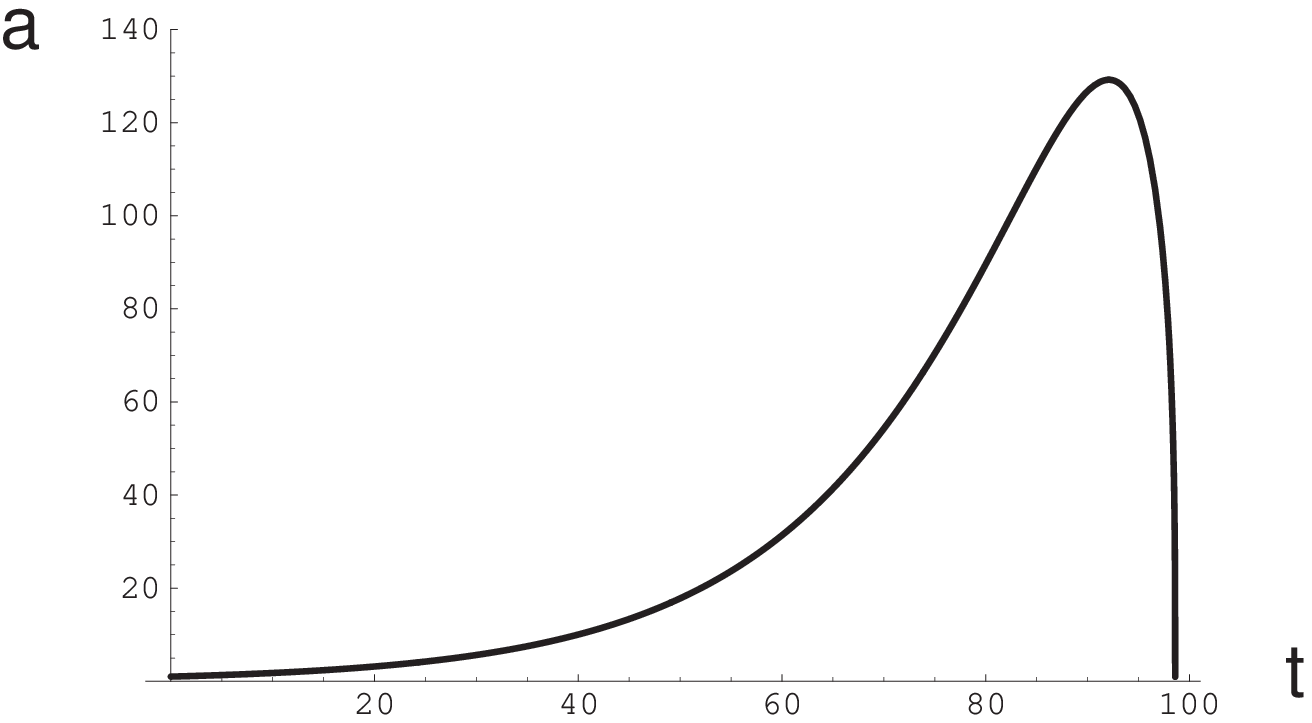} 
\caption{Evolution of the scalar field and the scale factor in the model with $V(\phi) = V_0 - {m^2\phi^2\over 2}$.  }
\label{conjunto4}
\end{figure}

The second model has the standard potential used for the description of spontaneous symmetry breaking, Eq. (\ref{SSB}), but we added to it a negative cosmological constant $\Lambda <0$ so that $V(\phi) = {\lambda\over 4} (\phi^2-v^2)^2 + \Lambda$ becomes negative in the minimum of $V(\phi)$ at $\phi = v$.   As we see in Fig. 2, in this case the scalar field experiences a stage of oscillations near the minimum of the effective potential with $V(\phi)= \Lambda <0$, but then it jumps off the minimum and blows up because of the ``negative friction'' in the collapsing universe. Interestingly, for most of the values of the parameters of the model (though not for all values), the field that originally moved towards the minimum with $\phi = +v$ blows up in the direction  $\phi \to -\infty$, and {\it vice versa}.  The reason is that at the initial stages of the development of the instability  the field $\phi$ is most efficiently accelerated by the negative friction if for a while it moves in a relatively flat direction, i.e.  from one minimum to another, instead of directly moving upwards.

\
 
\begin{figure}[h!]
\leavevmode
\centering \epsfysize=4cm
\includegraphics[scale=0.56]{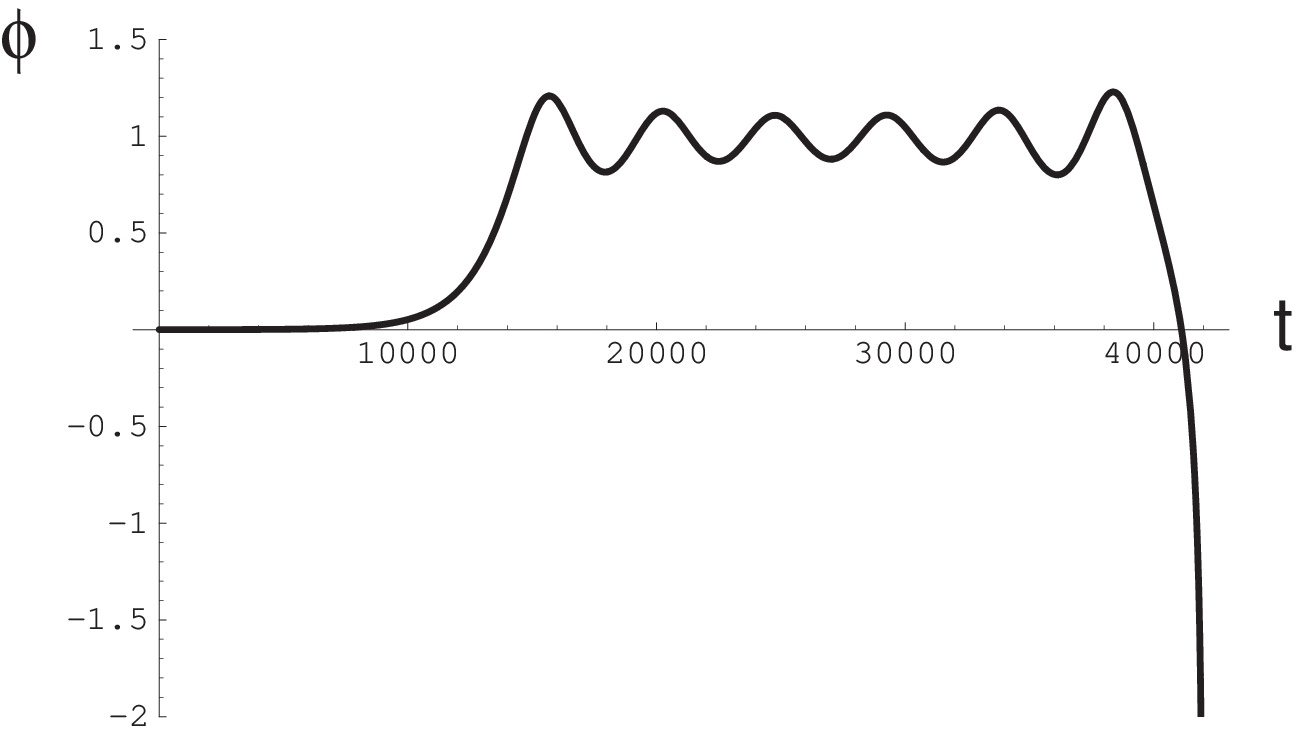}~~~
 \includegraphics[scale=0.56]{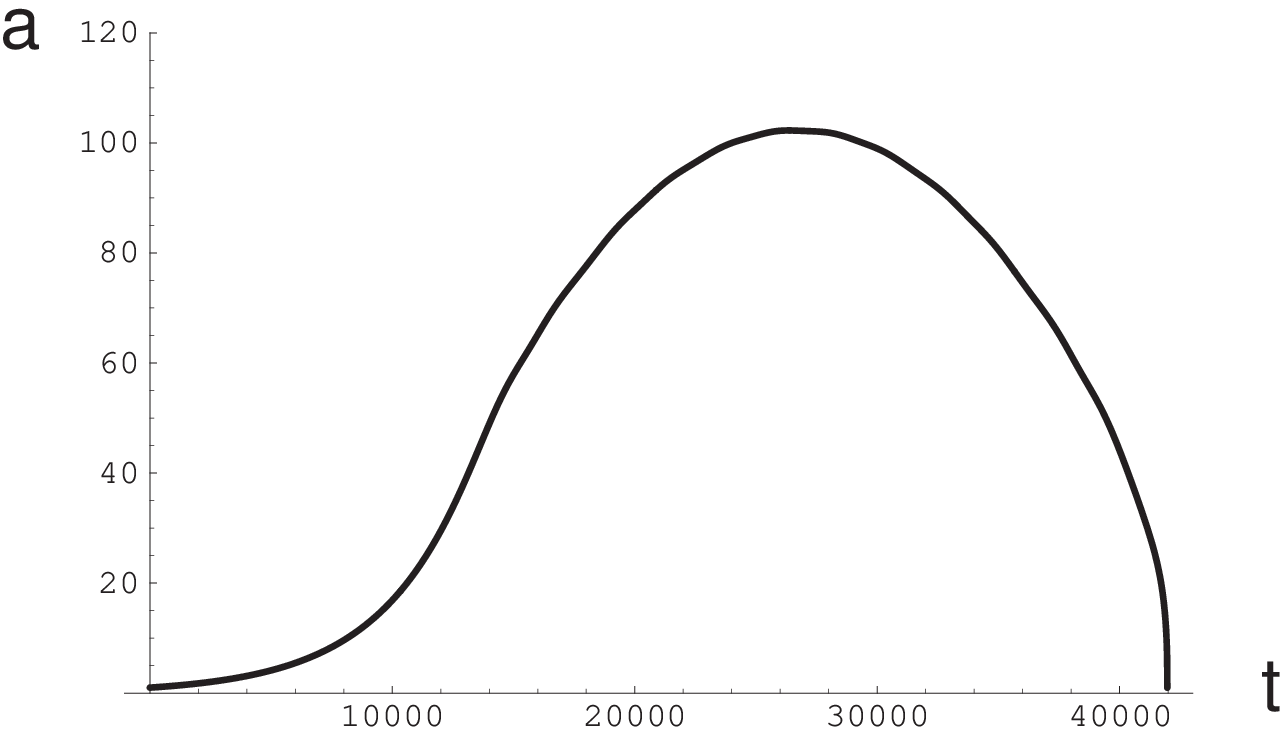} 
\caption{Evolution of the scalar field and the scale factor in the model with $V(\phi) = {\lambda\over 4} (\phi^2-v^2)^2 + \Lambda$, with $\Lambda <0$. }
\label{conjunto4a}
\end{figure}

However, when the field accelerates enough, it continues growing with the speed practically independent of $V(\phi)$. Consequently the structure of the singularity practically does not depend on the choice of the effective potential, unless the potential is exponentially steep. Indeed, let us assume for a second that in our investigation of the singularity one can neglect $V(\phi)$, and then check whether such an approximation is self-consistent.

In this approximation the energy density is determined not by $V(\phi)$ but by $\dot\phi^2/2$, and the corresponding equation of state is $p = \rho$. This leads to the collapse of the universe with $a(\tau) = \tau^{1/3}$, where $\tau$ is the time remaining before the singularity at $t = t_s$ is reached: ~$\tau = t_s-t$ \cite{book}. In this regime one has $H = 1/3\tau$, $H^2  =\dot\phi^2/6$, and  
\bea
|\phi| = \sqrt{2\over 3}\ \ln \tau, \qquad {\dot\phi^2\over 2} \sim {1\over 3\, \tau^2}\ .
\eea
It is clear therefore that for all  potentials $V(\phi)$ growing at large $\phi$ no faster than some power of $\phi$ one has $\dot\phi^2/2$
growing towards the singularity much faster than $V(\phi)$ (a power law singularity versus a logarithmic singularity). This means that one can indeed neglect $V(\phi)$ in the investigation of the singularity, essentially independently of the choice of the potential. The theories where   $V(\phi)$ becomes negative lead to a recollapsing flat universe with the energy density near the singularity being dominated by the kinetic energy of the  scalar field, and with $a(\tau) = \tau^{1/3}$.

Thus we see that when one takes gravity into account, then, from the point of view of the singular behavior of the field $\phi(t)$ and of the scalar factor $a(t)$, {\it the potentials having a global  minimum with $V(\phi) <0$ are as dangerous as the potentials unbounded from below.} The only difference is that in the theories where the potential is bounded from below there is an epoch where the scalar field oscillates near the minimum of $V(\phi)$ and the energy density of the universe (including the energy of particles produced by the decay of the field $\phi$) remains positive. If $\Lambda$ is sufficiently small, this epoch can be very long. On the other hand, the fast-roll inflation for small tachyonic mass $|m|$ in the theory with $V(\phi)$ unbounded from below also can be extremely long. In both cases we can live for a long time in an unstable state until the universe finally collapses.\footnote{It would be very interesting to find what may happen with the universe after the collapse, but despite many old and recent attempts to answer this question the problem of cosmological singularity still remains unresolved.}

Note that if one does not add a negative cosmological constant to the potential $V(\phi) = {\lambda\over 4} (\phi^2-v^2)^2$, or adds a positive cosmological constant, then the flat universe does not collapse but continues expanding for indefinitely long time. In the models studied in this paper the stage of accelerated expansion occurs in a state where $V(\phi)$ remains practically time-independent until the very end of the long stage of acceleration. Thus,  all available observational data can hardly tell the difference between the purely exponential expansion due to the positive cosmological constant and one of the regimes described above. This means, in particular, that one cannot say  whether we are going to end up in an empty exponentially expanding space, or the universe is going to slow down and approach Minkowski regime with zero cosmological constant, or it is going to collapse \cite{Krauss:1999br,Starobinsky:2000yw}. It is kind of sad that we cannot tell much about the way our part  of the universe is going to die. A partial encouragement is that  in many versions of inflationary theory the universe is capable of reproducing itself \cite{Vilenkin:1983xq,Eternal}. Moreover, instead of a simple mechanical self-replication, in the process of inflationary self-reproduction the universe can mutate, change its ``genetic code,'' and  re-emerge in all of its possible forms, giving rise to all possible types of life \cite{Eternal}.

\subsection{\label{early}Fast-roll inflation and scalar field perturbations in the very early universe}
After discussing the present stage of acceleration of the universe, it is time to return to the possibility of the fast-roll inflation in the very early universe, with $|m|$ just slightly below the Planck mass. As we have seen, in this case inflation is very short and inefficient. But we can find some use for it too.

Indeed, suppose that in addition to the field $\phi$ with $|m^2|  = O(H^2)$ there is another field $\chi$ that remains relatively light during the fast-roll inflation, $|m_\chi^2| \ll |m^2|  = O(H^2)$. Then the long-wavelength perturbations of this field are generated during inflation. If inflation occurs during the time $t$, the average amplitude of such perturbations is given by \cite{book}
\bea\label{fluct}
\sqrt{\langle\chi^2\rangle} = {H\over 2\pi}\sqrt{Ht} \ .
\eea
For example, using our estimates of the duration of inflation in the theory with the  potential $V(\phi)= V_0 - m^2\phi^2/2$ with $m = H$ (or, more realistically, in the theory $V(\phi) = {\lambda\over 4} (\phi^2-v^2)^2 = -{1\over 2} m^2\phi^2 + {m^2 \over 4 v^2} \phi^4 +{1\over 4} m^2 v^2$ with $m = H$) one finds, for $m = 1$,
\bea\label{fluct2}
\sqrt{\langle\chi^2\rangle} = {1\over 2\pi}\sqrt{7.5} \sim 0.4\ ,
\eea 
i.e. $\sqrt{\langle\chi^2\rangle}  \sim 10^{18}$ GeV.
Meanwhile for $m = H = 0.1$ one has 
\bea\label{fluct3}
\sqrt{\langle\chi^2\rangle} = {1\over 2\pi}\sqrt{15} \sim 0.06\ ,
\eea 
i.e. $\sqrt{\langle\chi^2\rangle}  \sim 10^{17}$ GeV.

This effect can be very useful for providing proper initial conditions for a subsequent stage of slow-roll inflation, if  the light field $\xi$ can play the role of the inflaton field.

There are some inflationary models, such as the simplest models of chaotic inflation with polynomial potentials, where inflation may start at a density very close to the Planck density and may continue until $V(\phi)$ becomes relatively small. (The last condition is necessary in order to produce small density perturbations, as well as gravitational waves with a small amplitude $\sim H/M_p$.) In such models inflation may begin and enter a stage of eternal self-reproduction if one has a single sufficiently homogeneous  domain of a smallest possible size $l = O(1)$ (i.e. $l \sim M_p^{-1}$) \cite{chaot,Eternal}.  Thus we believe that such models do not suffer from the problem of initial conditions \cite{book}.

However, there are many inflationary models where inflation  is possible only at $H$ many orders of magnitude smaller than $M_p$. For example, in most of the  models where inflation occurs near the top of the effective potential \cite{New}, as well as in most of the versions of hybrid inflation \cite{hybrid}, the Hubble constant practically does not change during inflation. In order to avoid excessive gravitational wave production in such models one must have $V \lesssim 10^{-10} M_p^4$ during inflation. Thus inflation in such models may start only rather late, at $t \sim H^{-1} \gg M_p^{-1}$. This leads to a problem of initial conditions required for inflation to begin in such models. In particular, it becomes difficult to explain the homogeneity of the universe on a scale $l \sim H^{-1} \gg M_p^{-1}$ at the beginning of inflation. If the universe is closed and dominated by nonrelativistic matter (or by an oscillating massive scalar field), then it collapses before it reaches the stage of inflation, unless its total mass is greater than $20M_p^2/H \gtrsim 10^6 M_p$ \cite{book}.

Fortunately, these problems disappear if one has an earlier  stage of ``bad'' inflation, which produces  unacceptable density perturbations and gravitational waves, but then triggers a subsequent stage of low-scale slow-roll inflation  \cite{Linde:1988yb}. 

The basic idea, in application to the fast-roll inflation, is as follows. One may have a  stage of a fast-roll inflation beginning at a nearly Planckian density, starting in a domain of a nearly Planckian size. The probability of such event should not be strongly suppressed \cite{book}. Inflation continues for a longer time if the universe initially was in a state with the scalar field very close to the top of the effective potential. There are at least three ways to achieve it.

First of all, this may happen due to the interaction of the inflaton field $\phi$ with other fields in the universe. These fields may be in a state of thermal equilibrium forcing the field $\phi$ to take the value $\phi = 0$, as in the new inflation scenario \cite{New}. It was very difficult to achieve this regime in the standard version of new inflation because we wanted to have the field $\phi$ extremely weakly interacting with matter  in order to produce small density perturbations. Such a field would not come to a state of thermal equilibrium with other fields  \cite{book}. Also,  we wanted inflation to begin at $t \sim H^{-1} \gg M_p^{-1}$, which was problematic; see a discussion above. In the case of the fast-roll inflation these requirements are not necessary and the corresponding complications do not appear.

Another possibility to put the field $\phi$ to the top of its effective potential is to use its interaction with other classical scalar fields, as in the hybrid inflation scenario \cite{hybrid,GuthRand,BLW}. Finally, one may consider a possibility of quantum creation of the universe from ``nothing.'' According to \cite{creation}, this process is not strongly suppressed if it occurs at the nearly Planckian energy density, and the universe is typically created at the top of the effective potential, which is exactly what we need to make inflation as long as possible.

Even in this case the stage of the fast-roll inflation may be very short, but it may be long enough to make the universe large and relatively homogeneous. For example,  after the short stage of inflation by a factor of $10^{3.3}$ in a model with $|m^2| = H^2$ (see Eq. (\ref{01a})) the total mass of a closed universe in the theory with $m = M_p$  becomes $10^{10} M_p$. For $m = H= 10^{-1} M_p$ the total mass of the universe after the fast-roll inflation becomes even greater, about $10^{15} M_p$. Thus there is no danger that it will collapse before $H$ drops down below $10^{-5} M_p$. 

Due to inflation and subsequent expansion of the universe, the fluctuations of the scalar field $\chi$ look like a nearly homogeneous classical scalar field $\chi$. However, this field takes different values in different parts of the universe. Those parts where the resulting values of $\chi$ correspond to good initial conditions for the slow-roll inflation driven by the field $\chi$ will expand exponentially, whereas the parts with bad initial conditions will remain small. In particular, in some parts of the universe quantum fluctuations may bring the field $\chi$ to the top of its effective potential, as in the new inflation scenario. The situation is especially interesting in the hybrid inflation scenario, where the fluctuations may grow only along the flat valley of the effective potential because the potential is very curved in all other directions. This is exactly what is needed to provide proper initial conditions for  hybrid inflation.  

Of course, this mechanism of generation of good initial conditions for the slow-roll inflation works even better if the initial stage of inflation is of the slow-roll type \cite{Linde:1988yb}. We just wanted to emphasize here that even a relatively short stage of a fast-roll inflation can resolve the problem of initial conditions for various models of the low-scale slow-roll inflation. 

In addition, even a short stage of the fast roll inflation may help us to solve the problem of symmetry breaking in SUSY GUTs and to justify a rather strong version of the anthropic principle.

Indeed, very soon after the invention of the new inflation scenario it became clear that inflation may consistently describe the universe consisting of different exponentially large parts with different laws of low-energy physics in each of them. This provided what, I believe, was the first  scientific justification of the weak anthropic principle \cite{Linde:1982gg}.  

One of the simplest implications of this idea was the discussion of symmetry breaking in the supersymmetric $SU(5)$ model. The problem was that this theory contains has a $SU(5)$ minimum with no symmetry breaking separated from the $SU(4) \times U(1)$  minimum and the $SU(3) \times SU(2)\times U(1)$ minimum by the barriers of the GUT height. Each of these minima have the same depth. If the universe initially were hot, then in the beginning the potential had only the $SU(5)$ minimum, so it was not clear how could we manage to wind up in the $SU(3) \times SU(2)\times U(1)$ minimum. 

A possible answer was that this happened because  inflation produced large perturbations of the field $\Phi$. (This is the field that breaks the $SU(5)$ symmetry.) In some parts of the universe this field drifted from the $SU(5)$ minimum to the $SU(4) \times U(1)$  minimum or the $SU(3) \times SU(2)\times U(1)$ minimum. As a result, the universe became divided into many exponentially large domains with different types of symmetry breaking, and we live in one of the exponentially large $SU(3) \times SU(2)\times U(1)$ domains  where life of our type is possible \cite{Linde:1983je}.

The only problem with this idea was that in the standard version of the new inflation scenario \cite{New} the value of $H^2$ is several orders of magnitude smaller than the curvature of the effective potential in the $SU(5)$ minimum. Therefore the required long-wavelength perturbations of the field $\Phi$ are not generated in the $SU(5)$ minimum. This problem disappears in the simplest versions of  chaotic inflation \cite{chaot}, where $H$ can be as large as $M_p$, i.e. it can easily exceed the GUT scale. That is why chaotic inflation \cite{chaot}, and especially eternal chaotic inflation \cite{Eternal}, where some parts of the universe may stay  for a long time in a state with $H$ approaching $M_p$, provided the best justification of the anthropic principle: The universe can probe all possible vacua even if they are separated by the barriers of nearly Planckian height.

In this respect the fast-roll inflation can provide an additional assistance: Even if the evolution of the universe does not begin in the slow-roll regime, very large fluctuations of the fields with masses $|m_\chi^2| \ll H^2$ may still be generated. This may divide the universe into large domains with different laws of low-energy physics, and then the size of each  domain may continue growing exponentially during a subsequent stage of a slow-roll inflation.

\subsection{Can we use the fast-roll inflation to solve the moduli and gravitino problems?}

It is most tempting to think about the possibility that a short stage of the fast-roll inflation at the electroweak scale or below it may solve the cosmological moduli and gravitino problems. Indeed, we have seen that the fast-roll inflation in the simplest model of spontaneous symmetry breaking with $v = 1$ and $m = 10^2$ GeV may lead to expansion of the universe by a factor of $10^7$, see Eq. (\ref{modulipr}). For $m < 10^2$ GeV the inflationary expansion is even much greater. Meanwhile for solving the gravitino and moduli problems in most cases it would be sufficient to have inflation by a factor of $10^4-10^5$ at $m,H \lesssim 10^2$ GeV, if the scalar field $\phi$ after inflation decays into ultra-relativistic particles. This would increase the total entropy of the universe by a factor of $10^{12}-10^{15}$. Typically this is sufficient for solving  the gravitino and moduli problems.

As we have seen, it is quite possible to obtain inflation by a factor of $10^4- {10^{5}}$  in the simplest models of the fast-roll inflation. And it is even easier to add few extra e-folds to the stage of thermal inflation  \cite{Thermal}, which was specifically invented to solve the gravitino and moduli problem. It was usually assumed that thermal  inflation ends as soon as the temperature drops down and the field begins to fall from the top of the effective potential. This assumption is often correct for the models with the scale of spontaneous symmetry breaking $v\ll 1$ \cite{Thermal}. However, as we have seen above, for $v \sim 1$ the universe may inflate by an additional factor $10^7$ after the field starts falling from the top of the effective potential.

This is a very encouraging result. However, whereas the fast-roll inflation followed by a prolonged stage of slow-roll inflation is a rather safe and interesting possibility, the models where inflation in the early universe ends up with a stage of a fast-roll inflation should be approached with some caution. The reason is that if the fast-roll inflation begins directly at the top of the effective potential, the adiabatic density perturbations produced at that stage may be too large.

\subsection{Adiabatic and isocurvature perturbations from the fast-roll inflation} 
Let us make a simple (and crude) estimate of the amplitude of the adiabatic density perturbations produced during the fast-roll inflation. According to \cite{pert},
\bea\label{fluct4}
 {\delta\rho\over\rho} \sim  {H\delta\phi\over \dot\phi} \ ,
\eea 
where $\delta\phi$ is the amplitude of the scalar field fluctuations generated during the time $H^{-1}$. For $|m^2| = O(H^2)$ one can take $\delta\phi \sim {H\over 2\pi}$, just like in the slow-roll inflation. For the initial value of the scalar field $\phi \sim m/2\pi \sim H/2\pi$, which corresponds to the initial level of the exponentially growing quantum fluctuations, one finds $\dot\phi \sim m^2/2\pi$, and ${\delta\rho\over\rho} = O(1)$.

Adiabatic perturbations produced at later stages of the fast-roll inflation have a much smaller amplitude (they have a very steep red spectrum), so typically they do not lead to any harmful  or desirable  consequences. But the consequences of the large perturbations produced at the very beginning of the process may be quite devastating  if they are not erased by a subsequent stage of the slow-roll inflation.  For example,  if density of the universe after inflation is dominated by nonrelativistic matter or by the energy of the oscillating scalar fields, these perturbations  can lead to a copious production of black holes of mass $M_{bh} \sim {M_p^2\over H}~ e^{3Ht_*}$.

This issue have been discussed for the first time in the context of a specific version of the hybrid inflation scenario proposed in \cite{GuthRand}. In this model it was assumed that after the standard stage of hybrid inflation, when the field moved slowly along a flat valley, there was a second stage of inflation during the process of spontaneous symmetry breaking. It was assumed that the amplitude of the symmetry breaking field is $O(M_p)$, and the curvature of the effective potential was of the same order as $H^2$. As a result, the second stage of inflation was short; it was very similar to the fast-roll inflation discussed in our paper. The main difference was that the model of Ref. \cite{GuthRand} was rather complicated; two fields were changing simultaneously during the second stage of inflation. In order to study this process it was necessary to use a specific combination of analytical and numerical methods \cite{BLW}. The result obtained in \cite{BLW} is that for the particular values of parameters taken in  \cite{GuthRand} the perturbations produced at the beginning of the second stage of inflation are too large, which leads to a significant  overproduction of  black holes. However, it is possible to change the parameters in such a way as to suppress density perturbations and avoid the problem of the black hole production \cite{BLW}.

The problem of black hole production does not arise at all if the fast-roll inflation is followed by a long stage of the slow-roll inflation, as discussed in Section \ref{early}. Similarly, this problem does not appear if the fast-roll inflation occurs at the present epoch, as described in Section \ref{now}. However, if one wants to have a fast-roll inflation at an intermediate scale in order to solve the moduli and gravitino problem, one should take this problem into account.

In addition to adiabatic perturbations with red spectrum sharply falling at large momenta, fast-roll inflation can produce isocurvature fluctuations with flat spectrum, related to the perturbations of light fields $\chi$. Historically, such perturbations were expected to be very small \cite{Brand}, but later it was shown that they can be as large and important as the adiabatic ones \cite{axions}. Isocurvature perturbations with flat spectrum usually do not lead to satisfactory cosmological consequences, but they can be converted to adiabatic ones if the fields $\chi$ decay to radiation  at the stage when they give a substantial contribution to the total energy density of the universe \cite{LW,MukhLin}. Thus in general one can use the fast-roll inflation as a source of adiabatic perturbations with flat spectrum, but this is a rather baroque possibility.

\section{Conclusions}

The standard inflationary paradigm is based on the idea of slow-roll inflation with $|V''| = |m^2| \ll H^2$. The slow-roll regime provides the simplest way to solve major cosmological problems and, simultaneously, to produce adiabatic perturbations with flat spectrum. But it may be useful to expand our toolbox a bit. It is quite possible that there was more than one inflationary stage. The first one happened in the very early universe, at the density that could be almost as large as the Planck density, $\rho \lesssim M_p^4$. We believe that the slow-roll condition should be satisfied at that stage. The second one apparently takes place right now, at a density that is $120$ orders of magnitude smaller than the Planck density.  As it was shown  above (see also \cite{KLPS}), this stage of inflation can be very long even if the slow-roll condition $|m^2| \ll H^2$ is strongly  violated, so that $|V''| = |m^2| \sim  10^4 H^2$!

A good thing about the fast-roll inflation is that the relation $|V''| = |m^2| = O(H^2)$  naturally appears in a large class of realistic models. Therefore it is quite possible that a stage of  the fast-roll inflation could occur not only at the present epoch, but also before or after the stage of the slow-roll inflation in the very early universe. This is a very interesting possibility that deserves further investigation.

\subsection*{Acknowledgments}
The author is grateful to G. Felder, and L. Kofman  for useful discussions. I am especially thankful to  R. Kallosh,  S.~Prokushkin and M.~Shmakova for the collaborative effort in our work on the fast-roll inflation in N=8 supergravity \cite{KLPS}. This work  was
supported by NSF grant PHY-9870115 and by the Templeton Foundation grant No. 938-COS273.

\


\begin{thebibliography}{99}


\bibitem{New} A.~D.~Linde,
``A New Inflationary Universe Scenario: A Possible Solution Of The Horizon, Flatness, Homogeneity, Isotropy And Primordial Monopole Problems,''
Phys.\ Lett.\ B {\bf 108}, 389 (1982); {\bf 116B}, 335, 340  (1982); A.~Albrecht and P.~J.~Steinhardt,
``Cosmology For Grand Unified Theories With Radiatively Induced Symmetry Breaking,''
Phys.\ Rev.\ Lett.\  {\bf 48}, 1220 (1982).

\bibitem{chaot} A.~D.~Linde,
``Chaotic Inflation,''
Phys.\ Lett.\ B {\bf 129}, 177 (1983).

\bibitem{hybrid}
A.~D.~Linde,
``Axions in inflationary cosmology,''
Phys.\ Lett.\ B {\bf 259}, 38 (1991);
A.~D.~Linde,
``Hybrid inflation,''
Phys.\ Rev.\ D {\bf 49}, 748 (1994)
[astro-ph/9307002].

\bibitem{book} A.D. Linde,  {\it  Particle  Physics  and
Inflationary Cosmology} (Harwood, Chur, Switzerland, 1990).

\bibitem{LythLiddle}  D. H. Lyth and A. R. Liddle, {\sl Cosmological Inflation
and Large-Scale Structure}, Cambridge University Press (2000).

\bibitem{pert} V.~F.~Mukhanov and G.~V.~Chibisov,
``Quantum Fluctuation And 'Nonsingular' Universe,''
JETP Lett.\  {\bf 33}, 532 (1981)
[Pisma Zh.\ Eksp.\ Teor.\ Fiz.\  {\bf 33}, 549 (1981)]; S.~W.~Hawking,
``The Development Of Irregularities In A Single Bubble Inflationary Universe,''
Phys.\ Lett.\ B {\bf 115}, 295 (1982); A.~A.~Starobinsky,
``Dynamics Of Phase Transition In The New Inflationary Universe Scenario And Generation Of Perturbations,''
Phys.\ Lett.\ B {\bf 117}, 175 (1982); A.~H.~Guth and S.~Y.~Pi,
``Fluctuations In The New Inflationary Universe,''
Phys.\ Rev.\ Lett.\  {\bf 49}, 1110 (1982); J.~M.~Bardeen, P.~J.~Steinhardt and M.~S.~Turner,
``Spontaneous Creation Of Almost Scale - Free Density Perturbations In An Inflationary Universe,''
Phys.\ Rev.\ D {\bf 28}, 679 (1983); V.~F.~Mukhanov,
``Gravitational Instability Of The Universe Filled With A Scalar Field,''
JETP Lett.\  {\bf 41}, 493 (1985)
[Pisma Zh.\ Eksp.\ Teor.\ Fiz.\  {\bf 41}, 402 (1985)];
V.~F.~Mukhanov, H.~A.~Feldman and R.~H.~Brandenberger,
``Theory of cosmological perturbations. Part 1. Classical perturbations. Part 2. Quantum theory of perturbations. Part 3. Extensions,''
Phys.\ Rept.\  {\bf 215}, 203 (1992).


        
        \bibitem{CMB} C.B. Netterfield {\it et al}, ``The BOOMERANG North America Instrument: a balloon-borne bolometric radiometer optimized for
measurements of cosmic background radiation
anisotropies from 0.3 to 4 degrees,'' astro-ph/0105148;
R. Stompor {\it et al}, ``Cosmological implications of the MAXIMA-I high resolution Cosmic Microwave Background anisotropy
measurement,'' astro-ph/0105062; N.W. Halverson {\it et al}, ``DASI First Results: A Measurement of the Cosmic
Microwave Background Angular Power Spectrum,'' astro-ph/0104489; P. de Bernardis {\it et al}, ``Multiple Peaks in the Angular Power Spectrum of the Cosmic Microwave Background: Significance and Consequences for Cosmology,'' astro-ph/0105296.

\bibitem{dinefisch} M.~Dine, W.~Fischler and D.~Nemeschansky,
``Solution Of The Entropy Crisis Of Supersymmetric Theories,''
Phys.\ Lett.\ B {\bf 136}, 169 (1984); G.~D.~Coughlan, R.~Holman, P.~Ramond and G.~G.~Ross, ``Supersymmetry And The Entropy Crisis,''
Phys.\ Lett.\ B {\bf 140}, 44 (1984).
 



\bibitem{KLPS} R.~Kallosh, A.~Linde, S.~Prokushkin and M.~Shmakova,
``Gauged Supergravities, de Sitter Space and Cosmology,''
[arXiv:hep-th/0110089].

\bibitem{FKL}
G.~Felder, J.~Garcia-Bellido, P.~B.~Greene, L.~Kofman, A.~D.~Linde and I.~Tkachev,
``Dynamics of symmetry breaking and tachyonic preheating,''
Phys.\ Rev.\ Lett.\  {\bf 87}, 011601 (2001)
[arXiv:hep-ph/0012142].
G.~Felder, L.~Kofman and A.~D.~Linde,
``Tachyonic instability and dynamics of spontaneous symmetry breaking,''
arXiv:hep-th/0106179. 


\bibitem{supernova} 
A.~G.~Riess {\it et al.}  [Supernova Search Team Collaboration],
``Observational Evidence from Supernovae for an Accelerating Universe
and a Cosmological Constant,''
Astron.\ J.\  {\bf 116}, 1009 (1998)
[astro-ph/9805201];
S.~Perlmutter {\it et al.}  [Supernova Cosmology Project Collaboration],
``Measurements of Omega and Lambda from 42 High-Redshift Supernovae,''
Astrophys.\ J.\  {\bf 517}, 565 (1999)
[astro-ph/9812133].

 






\bibitem{Hull}
C.~M.~Hull,
``A New Gauging Of N=8 Supergravity,''
Phys.\ Rev.\ D {\bf 30}, 760 (1984);
C.~M.~Hull,
``Noncompact Gaugings Of N=8 Supergravity,''
Phys.\ Lett.\ B {\bf 142}, 39 (1984).
C.~M.~Hull,
``The Construction Of New Gauged N=8 Supergravities,''
Physica {\bf 15D}, 230 (1985).
C.~M.~Hull,
``More Gaugings Of N=8 Supergravity,''
Phys.\ Lett.\ B {\bf 148}, 297 (1984).
C.~M.~Hull,
``The Minimal Couplings And Scalar Potentials Of The Gauged N=8 Supergravities,''
Class.\ Quant.\ Grav.\  {\bf 2}, 343 (1985);
C.~M.~Hull and N.~P.~Warner,
``Noncompact Gaugings From Higher Dimensions,''
Class.\ Quant.\ Grav.\  {\bf 5}, 1517 (1988).

\bibitem{Ahn:2001by}
C.~Ahn and K.~Woo,
``Domain wall and membrane flow from other gauged d = 4, n = 8  supergravity. I,''
arXiv:hep-th/0109010.


\bibitem{chaot2} A.~D.~Linde,
``Primordial Inflation Without Primordial Monopoles,''
Phys.\ Lett.\ B {\bf 132}, 317 (1983).


\bibitem{quint}
N.~Weiss,
``Possible Origins Of A Small Nonzero Cosmological Constant,''
Phys.\ Lett.\ B {\bf 197}, 42 (1987);~~~~
C.~Wetterich, ``Cosmology and Fate of Dilatation Symmetry'',
Nucl.\ Phys. \ B {\bf 302}, 668 (1988);
B.~Ratra and P.~J.~Peebles,
``Cosmological Consequences Of A Rolling Homogeneous Scalar Field,''
Phys.\ Rev.\ D {\bf 37}, 3406 (1988);~~~~
P.~G.~Ferreira and M.~Joyce,
 ``Structure formation with a self-tuning scalar field,''
Phys.\ Rev.\ Lett.\  {\bf 79}, 4740 (1997)
[astro-ph/9707286];
R.~R.~Caldwell, R.~Dave and P.~J.~Steinhardt,
``Cosmological Imprint of an Energy Component with General
Equation-of-State,''
Phys.\ Rev.\ Lett.\ {\bf 80}, 1582 (1998)
[astro-ph/9708069].


\bibitem{Hellerman:2001yi}
S.~Hellerman, N.~Kaloper and L.~Susskind,
``String theory and quintessence,''
JHEP {\bf 0106}, 003 (2001)
[hep-th/0104180];
W.~Fischler, A.~Kashani-Poor, R.~McNees and S.~Paban,
``The acceleration of the universe, a challenge for string theory,''
JHEP {\bf 0107}, 003 (2001)
[hep-th/0104181].

\bibitem{Halyo:2001fb}
E.~Halyo,
``Hybrid quintessence with an end or quintessence from branes and large  dimensions,''
[hep-ph/0105216].


\bibitem{Kallosh:2001tm}
R.~Kallosh,
``N = 2 supersymmetry and de Sitter space,''
arXiv:hep-th/0109168.


\bibitem{Krauss:1999br} L.~M.~Krauss and M.~S.~Turner,
``Geometry and destiny,''
Gen.\ Rel.\ Grav.\  {\bf 31}, 1453 (1999)
[arXiv:astro-ph/9904020].

\bibitem{Kaloper:1999tt}
N.~Kaloper and A.~D.~Linde,
``Cosmology vs. holography,''
Phys.\ Rev.\ D {\bf 60}, 103509 (1999)
[arXiv:hep-th/9904120].

\bibitem{Starobinsky:2000yw}
A.~A.~Starobinsky,
``Future and Origin of our Universe: Modern View,''
Grav.\ Cosmol.\  {\bf 6}, 157 (2000)
[arXiv:astro-ph/9912054].

\bibitem{Vilenkin:1983xq}
A.~Vilenkin,
``The Birth Of Inflationary Universes,''
Phys.\ Rev.\ D {\bf 27}, 2848 (1983).

\bibitem{Eternal}
A.~D.~Linde,
``Eternally Existing Self-reproducing Chaotic Inflationary Universe,''
Phys.\ Lett.\ B {\bf 175}, 395 (1986);
A.~D.~Linde,
``Particle Physics And Inflationary Cosmology,''
Physics\ Today {\bf 40}, 61 (1987);
A.~D.~Linde,
``The Self-reproducing inflationary universe,''
Scientific\ American\  {\bf 271}, 32 (1994);
A.~D.~Linde, D.~Linde and A.~Mezhlumian,
``From the Big Bang theory to the theory of a stationary universe,''
Phys.\ Rev.\ D {\bf 49}, 1783 (1994)
[arXiv:gr-qc/9306035];



\bibitem{Linde:1988yb}
A.~D.~Linde,
``Chaotic Inflation With Constrained Fields,''
Phys.\ Lett.\ B {\bf 202}, 194 (1988).


\bibitem{GuthRand} L.~Randall, M.~Soljacic and A.~H.~Guth,
``Supernatural Inflation: Inflation from Supersymmetry with No (Very) Small Parameters,''
Nucl.\ Phys.\ B {\bf 472}, 377 (1996)
[arXiv:hep-ph/9512439].



\bibitem{BLW} J.~Garcia-Bellido, A.~D.~Linde and D.~Wands,
``Density perturbations and black hole formation in hybrid inflation,''
Phys.\ Rev.\ D {\bf 54}, 6040 (1996)
[arXiv:astro-ph/9605094].

\bibitem{creation} A.~D.~Linde,
``Quantum Creation Of An Inflationary Universe,''
Sov.\ Phys.\ JETP {\bf 60}, 211 (1984)
[Zh.\ Eksp.\ Teor.\ Fiz.\  {\bf 87}, 369 (1984)];
Lett.\ Nuovo Cim.\  {\bf 39}, 401 (1984); A.~Vilenkin,
``Quantum Creation Of Universes,''
Phys.\ Rev.\ D {\bf 30}, 509 (1984).


\bibitem{Linde:1982gg}
A.~D.~Linde,
``The New Inflationary Universe Scenario,''
{\it  In:  Proceedings of the Nuffield Symposium ``The Very Early Universe,''  Cambridge 1982, 205-249}.

\bibitem{Linde:1983je}
A.~D.~Linde,
``Inflation Can Break Symmetry In Susy,''
Phys.\ Lett.\ B {\bf 131}, 330 (1983).


\bibitem{Thermal} D.~H.~Lyth and E.~D.~Stewart,
``Cosmology with a TeV mass GUT Higgs,''
Phys.\ Rev.\ Lett.\  {\bf 75}, 201 (1995)
[arXiv:hep-ph/9502417]; D.~H.~Lyth and E.~D.~Stewart,
``Thermal inflation and the moduli problem,''
Phys.\ Rev.\ D {\bf 53}, 1784 (1996)
[arXiv:hep-ph/9510204].




\bibitem{Brand}
M.~Axenides, R.~H.~Brandenberger and M.~Turner,
``Development Of Axion Perturbations In An Axion Dominated Universe,''
Phys.\ Lett.\ B {\bf 126}, 178 (1983).

\bibitem{axions}
A.~D.~Linde,
``Generation Of Isothermal Density Perturbations In The Inflationary Universe,''
JETP Lett.\  {\bf 40}, 1333 (1984)
[Pisma Zh.\ Eksp.\ Teor.\ Fiz.\  {\bf 40}, 496 (1984)];
A.~D.~Linde,
``Generation Of Isothermal Density Perturbations In The Inflationary Universe,''
Phys.\ Lett.\ B {\bf 158}, 375 (1985);
L.~A.~Kofman,
``What Initial Perturbations May Be Generated In Inflationary Cosmological Models,''
Phys.\ Lett.\ B {\bf 173}, 400 (1986).


\bibitem{LW} D.~H.~Lyth and D.~Wands,
``Generating the curvature perturbation without an inflaton,''
arXiv:hep-ph/0110002.


\bibitem{MukhLin} A.~D.~Linde and V.~Mukhanov,
``Nongaussian isocurvature perturbations from inflation,''
Phys.\ Rev.\ D {\bf 56}, 535 (1997)
[arXiv:astro-ph/9610219].


 






\end{thebibliography}
\end{document}